\documentclass[aps,reprint,nofootinbib]{revtex4-2}

\usepackage[english]{babel}
\usepackage[utf8]{inputenc}
\usepackage[normalem]{ulem}
\usepackage{amsmath,amssymb,amsfonts,mathrsfs,physics,bm,graphicx,xcolor,mathtools,dsfont,hyperref}
\usepackage{upgreek}
\hypersetup{
    colorlinks=true,
    linkcolor={red!40!black},
    citecolor={blue!50!black},
    urlcolor={blue!40!black}
    }
\graphicspath{{fig/}}
\definecolor{myred}{rgb}{0.85,0,0}
\definecolor{mygray}{rgb}{0.87, 0.87, 0.87}

\let\oldbibitem\bibitem 
\renewcommand{\bibitem}{
    \renewcommand{\doi}[1]{\texttt{\href{https://doi.org/##1}{doi:##1}}} 
    \let\bibitem\oldbibitem 
    \oldbibitem 
}

\newcommand{\exeter}{Department of Physics and Astronomy, University of Exeter, Stocker Road, Exeter EX4 4QL, United Kingdom}

\newcommand{\Deltak}{{\Delta}_{k}}
\newcommand{\Sk}{\mathrm{S}_{k}}
\newcommand{\bk}{b_k}
\newcommand{\rb}{{\rho}_\beta}
\newcommand{\rmix}{{\rho}_\mathrm{mix}}
\newcommand{\Trrb}{\Tr\big\{\rb^2\big\}}
\newcommand{\sansserif}[1]{{\fontfamily{cmss}\selectfont #1}}

\begin{document}

\title{The topology of data hides in quantum thermal states}

\author{Stefano Scali}
\email{s.scali@exeter.ac.uk}
\affiliation{\exeter}
\author{Chukwudubem Umeano}
\affiliation{\exeter}
\author{Oleksandr Kyriienko}
\affiliation{\exeter}

\begin{abstract}
We provide a quantum protocol to perform topological data analysis (TDA) via the distillation of quantum thermal states. Recent developments of quantum thermal state preparation algorithms reveal their characteristic scaling defined by properties of dissipative Lindbladians. This contrasts with protocols based on unitary evolution which have a scaling depending on the properties of the combinatorial Laplacian. To leverage quantum thermal state preparation algorithms, we translate quantum TDA from a real--time to an imaginary--time picture, shifting the paradigm from a unitary approach to a dissipative one. Starting from an initial state overlapping with the ground state of the system, one can dissipate its energy via channels unique to the dataset, naturally distilling its information. Therefore calculating Betti numbers translates into a purity estimation. Alternatively, this can be interpreted as the evaluation of the R\'{e}nyi 2--entropy, Uhlmann fidelity or Hilbert--Schmidt distance relative to thermal states with the embedded topology of simplicial complexes. Our work opens the field of TDA toward a more physical interpretation of the topology of data.
\end{abstract}

\maketitle


Extracting useful information from very large datasets~\cite{statista} is challenging given the tools available today, both from an algorithmic and a computational point of view. For this reason, approaches to analyze the crucial features of datasets have emerged. One of these approaches consists of extracting information from the ``shape'' of data, i.e. from its topological features, via the tools of topological data analysis (TDA)~\cite{Edelsbrunner_2002, Zomorodian_2004, CARLSSON_2005, Carlsson_2009}. TDA finds applications in several areas spanning physics~\cite{Kram_r_2016,Khasawneh_2016,Lee_2017,Pranav_2016,Tirelli_2021,Olsthoorn_2023}, medicine~\cite{Saggar_2018,Qaiser_2019}, and machine learning~\cite{Kovacev_Nikolic_2016,Hensel_2021}. Further applications can be found in Refs.~\cite{Wasserman_2018,Chazal_2021}. However, even TDA suffers from an unfavorable scaling with the system dimension. To bypass this problem, a natural extension comes in the form of quantum topological data analysis (QTDA). Since the first stages of QTDA~\cite{Lloyd2016}, these quantum algorithms have exploited the unitary evolution generated by the combinatorial Laplacian to extract its kernel information, key to access the topological properties. This requires state preparation based on Grover's search~\cite{grover1996fast} and quantum phase estimation (QPE)~\cite{kitaev1995quantum,Kitaev_1997}. Successive protocols proposed improved scaling by finding efficient representations of the combinatorial Laplacian while replacing Grover's search and QPE~\cite{Ubaru2021, Akhalwaya2022, Berry_2022}. Others found smart encoding strategies to provide, under certain conditions, an almost quintic advantage in space saving~\cite{mcardle2022streamlined}. Alternative protocols based on cohomology approaches~\cite{nghiem2023quantum} or hybrid quantum--classical pipelines focused on near--term devices~\cite{scali2023quantum} have been proposed.
While the existence of real instances of quantum advantage in QTDA is still debated~\cite{Berry_2022, schmidhuber2022complexitytheoretic}, the estimation of (normalized) Betti numbers on general chain complexes was shown to be \sansserif{DQC1}--hard, i.e. classically intractable~\cite{Gyurik_2022, cade2021complexity}. Safe from known protocols of dequantization~\cite{Tang_2019, gilyén2018quantuminspired, Chia_2020}, problems involving clique complexes have been shown to be $\mathsf{QMA}_1$--hard and contained in \sansserif{QMA}~\cite{crichigno2022clique, king2023promise}.

In this manuscript, we reinterpret topological data analysis from a quantum thermal state perspective. We propose a distinct paradigm to perform quantum topological data analysis, dubbed thermal--QTDA, that relies on the dissipative process defined by the combinatorial Laplacian associated with the dataset. The thermal state built from this process reveals, at low temperature, the topological features of the dataset. In this way, the evaluation of Betti numbers reduces to a purity test on the low--temperature quantum thermal state of the combinatorial Laplacian. We further interpret this result as the R\'{e}nyi 2--entropy of the thermal state and as the Uhlmann fidelity or the Hilbert--Schmidt distance between the maximally mixed state of the system and its imaginary time evolved version. Thermal--QTDA inherits the performance guarantees of the thermal state preparation protocol adopted and the efficiency of the purity test. This makes it a viable choice for early fault-tolerant quantum computers, with application in data analysis and machine learning.


\section{Betti numbers in quantum thermal states}
Let us briefly recall the theory behind QTDA and the evaluation of Betti numbers.
Consider a simplicial complex $\Gamma$ obtained from a dataset of dimension $N$, filtration distance, and a metric. Let $\Sk$ be the set of $k$--simplices of the complex $\Gamma = \{\Sk\}_{k=0}^{N-1}$. Let $\mathcal{H}_k$ be the $\binom{N}{k+1}$--dimensional Hilbert space spanned by all possible $k$--simplices. We refer to the single simplices $s_k\in\mathcal{H}_k$ with $s_k=j_0 \cdots j_k$ where $j_i$ is the $i$th vertex in $s_k$. Consider the boundary operator (map) $\partial_k : \mathcal{H}_k \mapsto \mathcal{H}_{k-1}$ defined by its action on the single simplices as $\partial_k|s_k\rangle = \sum_{l=0}^{k-1}(-1)^l|s_{k-1}(l)\rangle$, where $|s_{k-1}(l)\rangle = j_0 \cdots \hat{\jmath}_l \cdots j_k$ is the $(k-1)$--simplex obtained by removing the $l$th vertex from $s_k$. The action of the boundary operator on the $k$--simplices and their linear combinations determines the chain complex. Note that the operators and spaces just described can be restricted to the domain of the simplicial complex $\Gamma$ and it is usually indicated by a ``tilde'' as $\Tilde{\bullet}$. From the $k$--homology group of $\Gamma$ defined as $\mathbb{H}_k = \ker(\Tilde{\partial}_k)/\mathrm{im}(\Tilde{\partial}_{k+1})$, we obtain the Betti number as $\bk = \dim(\mathbb{H}_k)$. As a result of Hodge theory~\cite{lim2019hodge}, Betti numbers can also be evaluated as $\bk=\dim(\ker(\Deltak))=\dim(\Tilde{\mathcal{H}}_k)-\rank(\Deltak)$, where the combinatorial Laplacian $\Deltak$ relates to the boundary operators via $\Deltak = \Tilde{\partial}_k^\dagger\Tilde{\partial}_k + \Tilde{\partial}_{k+1}\Tilde{\partial}_{k+1}^\dagger$. From this expression, it is natural to look at spectral methods to estimate either the kernel or the rank of $\Deltak$. In this direction, since the seminal work by Lloyd~\cite{Lloyd2016}, several techniques relying on QPE or alternative spectrum evaluations have been developed~\cite{Ubaru2021, Akhalwaya2022, mcardle2022streamlined, Berry_2022, scali2023quantum}. In this work, we propose an alternative approach built upon a dissipative process or, equivalently, imaginary time evolution.

Given a dataset of dimension $N$, a filtration distance, and a metric, we can construct the simplicial complex $\Gamma = \{\Sk\}_{k=0}^{N-1}$ defining the topological properties of the dataset at that scale. Here, $\Sk$ is the set of the $k$--simplices in the complex. From this simplicial complex, we can construct the $k$th combinatorial Laplacian $\Deltak$ which encodes information of $\Gamma$ via its topological boundaries. We propose to estimate the $k$th Betti number $\bk$ of $\Gamma$ via the preparation of the low-temperature thermal state $\rb$ as
\begin{equation}
    \label{eq:betti}
    \bk = \lim_{\beta\rightarrow \infty} \Trrb^{-1} ,
\end{equation}
where $\beta=1/T$ is the inverse temperature and the thermal state $\rb$ is the Gibbs state $\rb = e^{-\beta\Deltak}/\Tr\big\{e^{-\beta\Deltak}\big\}$ (assuming Planck natural units). For Eq.~\eqref{eq:betti} to be valid, the thermalization process must start from an initial state overlapping with the ground state of the system. A possible choice is the maximally mixed state $\rmix = {\rho}_{\beta=0} = \mathds{I}_k/|\Sk|$ living in the Hilbert space $\Tilde{\mathcal{H}}_k$, spanned by the elements of $\Sk$. Note that the overlap of the initial state with the ground state (single or multiple) of the system affects the thermal state preparation, that is, the smaller the overlap, the slower the thermalization. Once an adequate initial state is formed, the thermalization process leads to the ground state of $\Deltak$, with the inverse purity of such state being its degeneracy. In case of degenerate ground state, the thermalization process will guide the system towards a thermal state that is an equiprobable mixture of the non--unique ground states. The overlap between this final state and the initial fully mixed state leads to the Betti numbers. This shows how the evaluation of Betti numbers via Eq.~\eqref{eq:betti} is related to the well--known expression $\bk=\dim(\ker(\Deltak))$~\cite{Lloyd2016}. In the following, we refer to the interpretation of QTDA via Eq.~\eqref{eq:betti} as thermal--QTDA.

\begin{figure}
    \centering
\includegraphics[width=\linewidth]{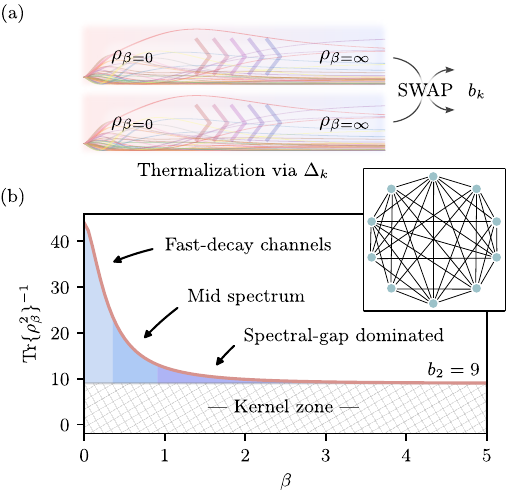}
    \caption{\textbf{Schematic of the protocol and Betti number evaluation.} In (a), two copies of the state $\rho_\beta$ are prepared at high temperature $\beta=0$. The states are thermalized to low temperature $\beta=\infty$ along the trajectories of dissipation dictated by the combinatorial Laplacian $\Deltak$. The resulting states are then swapped to perform a purity test and the Betti number evaluated through Eq.~\eqref{eq:betti}. In (b), we show an example numerical evaluation of the Betti number $b_2$ for the simplicial complex shown in the inset. We qualitatively identify regions of thermalization where specific decaying channels dominate. The hatched area represents the dimensionality of the kernel that remains inaccessible to the protocol.}
    \label{fig:betti_thermal}
\end{figure}

One may wonder what the physics behind Eq.~\eqref{eq:betti} is. Eq.~\eqref{eq:betti} is equivalent to evaluating the degeneracy of the ground state, relying on the fact that $\bk$ is defined as the kernel dimension of the $k$th homology group, as previously seen. To obtain this degeneracy, we start with an initial state with support on the ground state of the combinatorial Laplacian $\Deltak$, e.g. the maximally mixed state $\rmix$, and we cool it down to $\beta\rightarrow\infty$ via the channels of dissipation determined by $\Deltak$. In this limit, the state converges to the ground state of $\Deltak$. We show a schematic of the protocol in Fig.~\ref{fig:betti_thermal}(a). In Fig.~\ref{fig:betti_thermal}(b) we show an example of Betti number numerical evaluation via Eq.~\eqref{eq:betti} and the corresponding simplicial complex (inset). Notably, the thermalization approach is particularly suited to study ``the integer problem'' of Betti numbers. In fact, given the monotonic nature of Eq.~\eqref{eq:betti}, we can approximate the Betti number as $\lfloor b_k \rfloor$ when the state is prepared at large but finite $\beta$. This suggests the possibility of using an approximate thermal state preparation to reach a low--enough finite temperature.


\section{Interpretations}
Eq.~\eqref{eq:betti} can be interpreted as the Uhlmann fidelity between the maximally mixed state $\rmix$ and its imaginary--time evolved version $\rmix(\tau)$. This becomes possible if one translates the unitary evolution problem from Minkowski space to Euclidean space by allowing time to take imaginary values and replacing $t = -i\tau$. Thus, the Uhlmann fidelity can be expressed as
\begin{align}
    \label{eq:fidelity}
    \mathscr{F}(\rmix, \rmix(\tau)) &= \Tr\big\{(\rmix^{1/2} \rmix(\tau) \rmix^{1/2})^{1/2}\big\}^2 \nonumber \\
    &= \Tr\big\{e^{-\Deltak \tau}\big\}^2 / (\Tr\big\{e^{-2\Deltak \tau}\big\} |\Sk|) \nonumber \\
    &= \Trrb^{-1} / |\Sk| ,
\end{align}
with $\rmix(\tau) = e^{-\Deltak \tau} \rmix e^{-\Deltak \tau} / \big(\Tr\big\{\rmix e^{-2 \Deltak \tau}\big\}\big)$, where the factor at the denominator makes sure the density matrix is normalized throughout the evolution. While the components populating the kernel contribute to the fidelity at all imaginary times, the remaining components decay to zero exponentially. We obtain the final line in Eq.~\eqref{eq:fidelity} by means of Wick rotation, i.e. replacing $\tau = \beta$, thus reinterpreting the imaginary time as a temperature. These two pictures, imaginary time in quantum mechanics and temperature in statistical physics, are indeed formally related through analytic continuation~\cite{zee2010quantum} by the Osterwalder–Schrader theorem~\cite{Osterwalder_1973, Osterwalder_1975}.

As a natural extension, Eq.~\eqref{eq:betti} can also be interpreted as the quantum version~\cite{Petz_1986,Muller_Lennert_2013} of the R\'{e}nyi 2--entropy~\cite{renyi1961measures} of the thermal state $\rb$,
\begin{equation}
    \label{eq:renyi}
    \mathscr{H}_2(\rb) = \log\big(\Trrb^{-1}\big) .
\end{equation}
Here, the base of the logarithm determines the unit of information. The R\'{e}nyi 2--entropy is often referred to as the collision entropy. R\'{e}nyi entropies are of crucial interest for the estimation of the statistical properties of quantum states, finding applications as entanglement measures~\cite{Wang_2016,Elben_2018,Brydges_2019,Elben_2019}, in the estimation of Gaussianity of quantum states~\cite{Adesso_2012,Park_2021}, and as a measure of non--stabilizerness (also known as magic)~\cite{Leone_2022}. Given their ubiquity, several quantum algorithms have been developed to estimate R\'{e}nyi entropies~\cite{Li_2019,Acharya_2020,Subramanian_2021,Wang_2023}.

Finally, Eq.~\eqref{eq:betti} can be interpreted in terms of the Hilbert--Schmidt distance (Schatten 2--norm),
\begin{align}
    \label{eq:distance}
    \mathscr{D}_{\mathrm{HS}}(\rmix,\rb) &= ||\rmix - \rb||_2^2 \nonumber \\
    &= \Trrb - (2|\Sk|-1)/|\Sk|^2 .
\end{align}
This distance is often employed as the cost function in variational quantum algorithms (VQAs)~\cite{LaRose_2019,Arrasmith_2019,Khatri_2019,Tan_2021,Ezzell_2023}.


\section{Routines}
Assuming that we are given two copies of the thermal state $\rb$ at low temperature, we only need to perform a purity test to implement Eq.~\eqref{eq:betti} on a quantum computer. While this can be done in the form of a traditional SWAP test or its destructive variant~\cite{Garcia_Escartin_2013}, we note that some generalizations with short--depth circuit have been proposed~\cite{Johri_2017,Cincio_2018,Suba__2019}. Alternatively, purity can be evaluated via single distinct classical shadows of the thermal state~\cite{Elben_2018, Brydges_2019, Elben_2019, Elben_2022}. This approach trades two identical copies of the thermal state and a simple measurement routine for a single copy of the thermal state and a larger number of (randomized) measurements. In the following, we consider a simple SWAP test. To perform it, we need an additional ancilla qubit. We place a Hadamard gate on the ancilla, a Fredkin (CSWAP) gate controlled on the ancilla between the two copies of qubits in the state registers, and then again a Hadamard gate on the ancilla. By measuring the ancilla qubit multiple times, we construct the probabilities $P_0$ and $P_1$ relative to the outcome 0 and 1 respectively. We obtain the purity of the thermal state from $\Trrb = P_0-P_1$~\cite{Kobayashi_2003}. Once the SWAP test has been performed, the result is obtained by truncating (flooring) the Betti number in Eq.~\eqref{eq:betti}, $\lfloor b_k \rfloor$, as previously mentioned.

The thermal state can be prepared in several ways. A nature--inspired thermal state preparation can be found in Ref.~\cite{chen2023quantum} where the authors propose a protocol to simulate the Lindbladian (or the relative discriminant proxy) whose fixed point is approximately a quantum Gibbs state. By means of this construction and the introduction of an operator Fourier transform (FT) for the Lindblad operators, the authors give the recipes for incoherent and coherent implementations. See later for an example implementation using such coherent approach.
Several variational quantum algorithms have been developed to prepare thermal states using imaginary time evolution~\cite{McArdle_2019}, open system dynamics~\cite{Endo_2020}, and hybrid quantum circuits using classical neural networks~\cite{Liu_2021}.
Thermal--QTDA also opens the door to possible quantum--inspired implementations based on thermal tensor network (TTN) states. Examples of these are the realization of thermal states using minimally entangled typical thermal states (METTS) and imaginary time evolution~\cite{Motta_2019} or the exponential tensor renormalization group (XTRG)~\cite{Chen_2018} producing accurate low--temperature thermal states by exponentially evolving the matrix product operator (MPO) along a path of imaginary time evolution.
In Ref.~\cite{Coopmans_2023}, an interesting approach to the estimation of linear functions of $\rb$ avoids the direct construction of the mixed state $\rb$. The authors combine pure thermal quantum states~\cite{Sugiura_2012,Sugiura_2013} and classical shadow tomography~\cite{Ohliger_2013,Aaronson_2018,Huang_2020} to estimate several Gibbs state expectation values. However, thermal--QTDA requires the estimation of nonlinear functions of $\rb$. In this direction, the authors envision improvements to their algorithm in the form of derandomization~\cite{Huang_2021,Zhang_2021,Hadfield_2022}.

\section{Runtime}
The cost of thermal--QTDA is inherited from the scaling of the two subroutines, the quantum thermal state preparation and the purity test. For the latter, the SWAP test requires two copies of the quantum thermal state and an ancilla qubit, with a final measurement on the ancilla qubit. The destructive alternative requires two copies of the quantum thermal state but trades the additional ancilla with a final measurement onto the full set of qubits. A test result with additive error $\epsilon$ requires $\mathcal{O}(\epsilon^{-2})$ runs.

Now, the quantum thermal state preparation. When estimating Betti numbers, quantum phase estimation explicitly introduces an inverse linear dependence on the spectral gap of the combinatorial Laplacian, $\mathcal{O}(N^3\delta_\mathrm{gap}^{-1})$~\cite{Lloyd2016}. In contrast, the cost of estimating Betti numbers via quantum thermal state preparation varies depending on the chosen algorithm. For example, simulating an incoherent version of the Lindbladian via the operator FT costs $\Tilde{\mathcal{O}}(\beta t_\mathrm{mix}^2/\epsilon)$ while the coherent version via the discriminant proxy and an adiabatic path to low temperature costs $\Tilde{\mathcal{O}}\big((\epsilon\beta^2\norm{H} + \beta)/\epsilon\lambda_\mathrm{gap}^{3/2}\big)$~\cite{chen2023quantum}. Here, we use the soft-O notation $\Tilde{\mathcal{O}}$, which corresponds to scaling that ignores constant and logarithmic factors. The mixing time of the Lindbladian, $t_\mathrm{mix}$, and the minimum spectral gap of the discriminant proxies along the adiabatic path, $\lambda_\mathrm{gap}$, are related by the approximate detailed balance introduced in Ref.~\cite{chen2023quantum}. In the perspective of open systems and Gibbs samplers, the spectral gap of the Lindbladian is also related to the spectral gap of the combinatorial Laplacian. However, this relation strongly depends on the characteristics of the simplicial complex (system Hamiltonian) considered~\cite{Kastoryano_2013, Temme_2013, Kastroyano_2014, Capel2020, chen2021, Wild_2021}. As a general guideline, a faster convergence to the thermal state can be obtained in the presence of stronger dissipative coupling, ad--hoc transitions boosting convenient channels of dissipation, or higher finite temperatures of thermalization. Note that, producing thermal states for long mixing times or closing spectral gaps can be challenging. Hard instances of thermal state preparation are expected regardless of the algorithm chosen, since the ground state preparation problem is \sansserif{QMA}--hard~\cite{Kitaev_2002,kempe2003,Kempe_2004}.

\begin{figure}
    \centering
    \includegraphics[width=\linewidth]{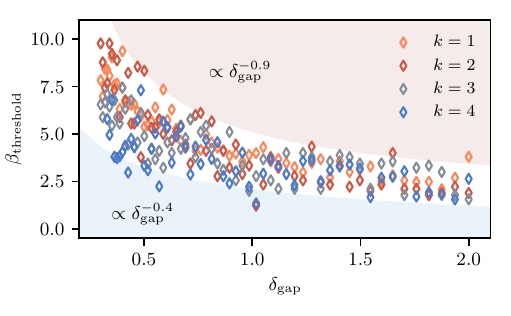}
    \caption{\textbf{Spectral gap scaling.} We sample the imaginary time/inverse temperature needed to obtain $\frac{\dd}{\dd\tau}\Tr\big\{\rmix e^{-\Deltak \tau}\big\}\leq10^{-3}$ for random instances of simplicial complexes with $N=10$ and $k=1,2,3,4$. We find a corresponding inverse temperature dependence on the spectral gap of the combinatorial Laplacian of $\beta_\mathrm{threshold} = \mathcal{O}(\delta_\mathrm{gap}^{-(0.4\divisionsymbol 0.9)})$. Here, the notation $a\divisionsymbol b$ indicates a range of values between $a$ and $b$.}
    \label{fig:scaling}
\end{figure}

In Fig.~\ref{fig:scaling} we show the spectral gap scaling of thermal--QTDA numerically evaluated for random simplicial complexes chosen with $N=10$ and $k=1,2,3,4$. The initial state $\rmix$ is thermalized until the stopping criterion is met, that is, $\frac{\dd}{\dd\tau}\Tr\big\{\rmix e^{-\Deltak \tau}\big\}\leq10^{-3}$. We refer to the smallest inverse temperature that satisfies such criterion as the threshold inverse temperature $\beta_\mathrm{threshold}$. We find this to be $\beta_\mathrm{threshold} = \mathcal{O}(\delta_\mathrm{gap}^{-(0.4\divisionsymbol 0.9)})$, where the notation $a\divisionsymbol b$ indicates a range of values between $a$ and $b$, suggesting an inverse--sub--linear dependence of thermal--QTDA on the spectral gap of the combinatorial Laplacian.


\begin{figure*}
    \centering
    \includegraphics[width=\linewidth]{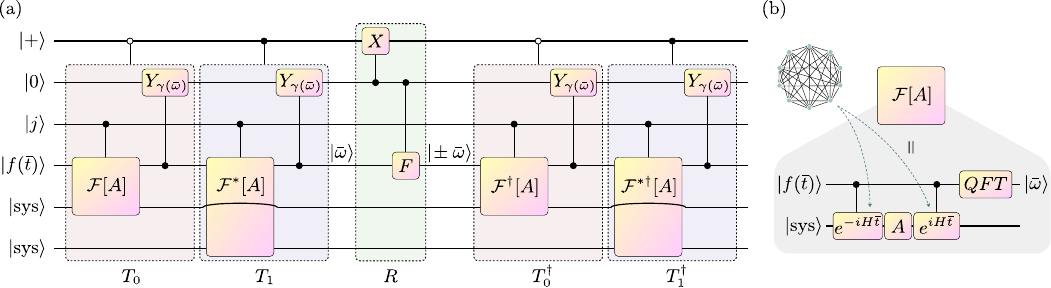}
    \caption{\textbf{Discriminant proxy and operator Fourier transform circuits.} In (a) we show a circuit for preparing quantum thermal states $\rb$ as a sub--routine of thermal--QTDA. These are prepared as the top eigenvector of the discriminant proxy $D_\beta$ via quantum simulated annealing~\cite{chen2023quantum}. These copies are then used to estimate Betti numbers via a SWAP test and Eq.~\eqref{eq:betti}. The operator Fourier transform $\mathcal{F}[A]$ from the discriminant proxy implementation is shown in (b). This routine encodes the information of the simplicial complex $\Gamma$ (and the relative combinatorial Laplacian $\Deltak$) into the jump operators $A$.}
    \label{fig:discriminant}
\end{figure*}

\section{Example implementation}
In Fig.~\ref{fig:discriminant}, we show an example circuit to prepare the thermal state $\rb$. Following the coherent approach in Ref.~\cite{chen2023quantum}, we implement the discriminant proxy $D_\beta$ governing the dissipation dictated by the combinatorial Laplacian,
\begin{align}
    \label{eq:discriminant}
    D_\beta &\coloneqq \frac{1}{|\mathcal{A}|} \sum_{j, \Bar{\omega}} \sqrt{\gamma(\Bar{\omega})\gamma(-\Bar{\omega})} A_j(\Bar{\omega}) \otimes A_j^{*}(\Bar{\omega}) \nonumber\\
    &-\frac{\gamma(\Bar{\omega})}{2} \Big(A_j^{\dagger}(\Bar{\omega})A_j(\Bar{\omega})\otimes\mathds{I} + \mathds{I}\otimes A_j^{*\dagger}(\Bar{\omega}) A_j^{*}(\Bar{\omega})\Big) .
\end{align}
The top eigenvector of the discriminant proxy $D_\beta$ is the canonical purification of the thermal state at inverse temperature $\beta/2$, that is $|\sqrt{\rb}\rangle \propto \sum_i e^{-\beta E_i / 2} |\psi_i\rangle \otimes |\psi_i\rangle$, where the $E_i$ are the eigenvalues of the Hamiltonian operator $H=\Deltak$. By using two copies of this purified version of the thermal state, we can perform the SWAP test and evaluate $\Trrb$, getting Betti numbers via Eq.~\eqref{eq:betti}.

In Eq.~\eqref{eq:discriminant}, $A_j(\Bar{\omega})$ are the Fourier--transformed versions of the Heisenberg time--evolved jump operators $A_j(\Bar{t})$ (see later), $\gamma(\Bar{\omega})$ are the jump weights, and $\mathcal{A}$ is the set of jump operators $A_j$. For convenience, we choose the $A_j$ to be the self--adjoint single--site Pauli operators. We also choose the transition weights $\gamma(\Bar{\omega})$ to be the Metropolis weights. This choice ensures that the transition weights satisfy $\gamma(\Bar{\omega})/\gamma(-\Bar{\omega})=e^{\beta\Bar{\omega}}$, consequently the detailed balance condition. The ``bar'' indicates discretized frequencies in the set of frequencies $S_{\omega_0}$. These are multiples of a base frequency $\omega_0$, which is defined by the relation $\omega_0t_0=2\pi/M$, where $t_0$ is the base discretized time and $M$ is the number of discretized points. To cover the full set of transition (Bohr) frequencies defined by $H$, the number of points and the base frequency are chosen such that $M\omega_0/2\geq\norm{H}$.

To build the circuit implementing Eq.~\eqref{eq:discriminant}, we rewrite the discriminant in terms of the reflection $R$ and the isometry $T'$ as $\mathds{I}+{D}_\beta = {T}'^\dagger {R} {T}'$~\cite{Szegedy,wocjan2021szegedy,chen2023quantum}. The isometry ${T}'$ is a coherent sum of the two sub--isometries ${T}_0$ and ${T}_1$, ${T}'=|0\rangle\otimes{T}_0 + {T}_1\otimes|1\rangle$. Each of these sub--isometries returns a superposition of jump operators applied on the system register $|\mathrm{sys}\rangle$ of interest. This is done via the operator FT $\mathcal{F}[A]$ controlled on the ancillary register $|j\rangle$ that carries the weight distribution of the jump operators. In ${T}_0$ and ${T}_1$, we also find the encoder for the transition rates, ${Y_{\gamma(\Bar{\omega})}}$. ${R}$ is a reflection (${R}^2=\mathds{I}$) that implements a bit flip $X$ on the ancilla qubit $|+\rangle$ dedicated to the control of the sub--isometries ${T}_0$ and ${T}_1$ and a sign change to the Bohr frequency register. These operations are all controlled by the ancillary qubit register ($|0\rangle$) storing the transition rates. If the reflection is triggered, it generates a cross term between the two $|\mathrm{sys}\rangle$ copies.

The operator FT $\mathcal{F}[A]$, introduced in Ref.~\cite{chen2023quantum}, implements the following weighted transform on the jump operators, $A(\Bar{\omega}) := \sum_{\Bar{t}}e^{-i\Bar{\omega}\Bar{t}}f(\Bar{t})A(\Bar{t})$, where $A(\Bar{t})=e^{iH\Bar{t}}Ae^{-iH\Bar{t}}$ are the time evolved versions of the jump operators. Here, the unique channels of dissipation defined by the combinatorial Laplacian $\Deltak$ enter into the picture. By equating $H$ with $\Deltak$ as previously stated, we build jumps that encode information of the combinatorial Laplacian and that dissipate the energy of the thermal state accordingly. We show the circuit implementing $\mathcal{F}[A]$ in Fig.~\ref{fig:discriminant}(b). Given the weighted nature of the FT implemented, the time evolution of the jump operators is controlled by the weighting function $f(\Bar{t})$. After the control, the weights are quantum Fourier transformed into the Bohr frequencies $|\Bar{\omega}\rangle$ to then control the transition rates $\gamma(\Bar{\omega})$ (Boltzmann weights). Notably, when the weighting function follows a Gaussian distribution, the sub--routine achieves the performance scaling of boosted phase estimation~\cite{chen2023quantum, nagaj2009fast}. Then, the unitary ${Y_{\gamma(\Bar{\omega})}}$ encodes the transition rates $\gamma(\Bar{\omega})$ in the amplitudes of the ancillary qubit $|0\rangle$. The unitary is controlled by the output of the operator FT $\mathcal{F}[A]$, namely the Bohr frequency read--out $|\Bar{\omega}\rangle$. Finally, $F$ is the negation of the Bohr frequency register. This makes sure that the ensuing isometries ${T}_0^\dagger$ and ${T}_1^\dagger$ produce the correct jump operator, that is $A(\Bar{\omega}) = A^{\dagger}(-\Bar{\omega})$ (see details in Ref.~\cite{chen2023quantum}).

To finally find the top eigenvector of the discriminant proxy, i.e. the quantum thermal state, we need to perform quantum simulated annealing. Having access to the block encoding of $D_\beta$, we can build a unitary $U_{D_\beta}$ with $\mathcal{O}(\lambda_\mathrm{gap}^{-1/2}(D_\beta))$ calls of $D_\beta$. Starting from the maximally mixed state at very high temperature $\beta_0\approx0$, we proceed with adiabatic steps of unitaries through a path of decreasing temperatures, $U_{D_{\beta_0}} \rightarrow \cdots \rightarrow U_{D_{\beta_\mathrm{target}}}$, finally reaching a very low temperature $\beta_\mathrm{target}$. At the end of this path, we obtain the top eigenvector of the discriminant, thus the thermal state. Once we have prepared two copies of the state, we can perform the SWAP test and extract the Betti numbers.


\section{Conclusions and outlook}
In this manuscript, we reinterpreted a pillar of quantum topological data analysis, that is, the estimation of Betti numbers, under the lens of quantum thermal states. We dubbed this interpretation thermal--QTDA. The protocol relies upon the preparation of two copies of quantum thermal states at low temperature and a purity test. The thermalization of the states happens through the unique channels of dissipation of the combinatorial Laplacian. The purity test is naturally related to other quantities of interest such as the R\'{e}nyi 2--entropy, Uhlmann fidelity or Hilbert--Schmidt distance. The scaling of thermal--QTDA routines directly depends on the scaling of its two sub--routines, the thermal state preparation and the purity test. We presented a possible coherent Lindbladian approach for the former, while a simple SWAP test is sufficient for the latter.

Thermal--QTDA has theoretical and practical importance. On the theoretical side, it hints to possible quantum thermodynamic quantities as the tools of interest to perform QTDA tasks. Here, one may wonder if thermal states prepared from simplicial complexes can reveal additional topological features of the data. On the practical side, it opens the field of QTDA to an entire new set of possible implementations via quantum thermal state preparation protocols. Here, the crucial step is breaking away from the traditional scaling of quantum phase estimation, which may lead to an advantageous dependence on the spectral gap of the combinatorial Laplacian. From a perspective of applications, extensions of our approach to the evaluation of persistent Betti numbers are desirable. In fact, these enclose additional information on the persistence of features and can yield enhanced explainable AI protocols~\cite{neumann2019limitations}. We expect advances in QTDA will facilitate the use of topological properties as features of datasets, aiding the building of models with high predictive power and excellent generalization. 


\section{Acknowledgements}
The authors thank Ravindra Mutyamsetty, Zhihao Lan and Okello Ketley for useful discussions. The authors acknowledge the support from the Innovate UK ISCF Feasibility Study project number 10030953 granted under the ``Commercialising Quantum Technologies'' competition round 3.


\bibliographystyle{apsrev4-2}
\bibliography{main}

\end{document}